\documentclass[twocolumn,showpacs,preprintnumbers,amsmath,amssymb]{revtex4}

\usepackage{graphicx}
\usepackage{dcolumn}
\usepackage{bm}

\begin{document}

\title{Adaptive molecular resolution via a continuous change of the
  phase space dimensionality}
\author{Matej Praprotnik}
\altaffiliation{On leave from the National Institute of Chemistry, Hajdrihova 19,
                 SI-1001 Ljubljana, Slovenia. Electronic Mail: praprot@cmm.ki.si}
\author{Kurt Kremer}
\author{Luigi Delle Site}
\affiliation{%
Max-Planck-Institut f\"ur Polymerforschung, Ackermannweg 10, D-55128 Mainz, Germany
}%

\begin{abstract}
For the study of complex synthetic and biological molecular
systems by computer simulations one is still restricted to simple
model systems or to by far too small time scales. To overcome this
problem multiscale techniques are being developed for many
applications. However in almost all cases, the regions of
different resolution are fixed and not in a true equilibrium with
each other. We here give the theoretical framework for an
efficient and flexible coupling of the different regimes. The
approach leads to an analog of a geometry induced phase transition
and a counterpart of the equipartition theorem for fractional
degrees of freedom. This provides a rather general formal basis
for advanced computer simulation methods applying different levels
of resolution.
\end{abstract}
\maketitle
A long standing and often most challenging problem in condensed
matter physics, up to some simple crystalline materials, is to
understand the microscopic origin of macroscopic properties. While
in certain cases a microscopic scale can be clearly separated from
a macroscopic one, this is not the case for most experimental
systems -- details of the local interaction and generic/universal
aspects are closely related. Already the  proper determination of
the fracture energy and crack propagation in crystalline materials
requires a hierarchical and interrelated description, which links
the breaking of the interatomic bonding in the fracture region to
the response of the rest of the system on a micron scale
\cite{Rottler:2002}. The presence of microscopic chemical
impurities in many metals and alloys changes their macroscopic
mechanical behavior \cite{Jiang:2004,Lu:2005}. Even more
complicated are synthetic and biological soft matter systems.
Whether one is dealing with the morphology, the glass
transition of a polymeric system, the function of a molecular
assembly, e.g. for electronic applications or studies
ligand-protein recognition or protein-protein interaction, in all
cases the generic soft matter properties, such as matrix or chain
conformation fluctuations, and details of the local chemistry
apply to roughly the same length scales. For all these problems,
which due to their complexity are heavily studied by computer
simulations, there is a common underlying physical scenario: the
number of degrees of freedom (DOFs) involved is very large and the
exhaustive exploration of the related phase space is prohibitive.
For many questions, however, such a deep level of detail in the
description is only required locally.

Theoretical methods employed to study these systems span from
quantum-mechanical to macroscopic statistical approaches. Their
efficiency and scope increases significantly if two or more such
different approaches are combined into hybrid multiscale schemes.
This is the case for the quantum based QM/MM approach
\cite{Laio:2002} and that of dual scale resolution techniques
\cite{DelleSite:2002,DelleSite:2004,Villa:2004,Neri:2005,Rafii:1998,
Broughton:1999,Smirnova:1999,Csanyi:2004,OConnell:1995,Hadjicostantinou:1999,
Li:1998,Flekkoy:2000, Delgado:2003, Baschnagel:2000} aiming at bridging the atomistic and mesoscale length
scale. However, the common feature and limitation of all these
methods is the fact that the regions or parts of the system
treated at different level of resolution are fixed and do not
allow for free exchange. It is exactly this constraint one has to
overcome in order to study typical complex, fluctuating molecular
systems with a higher computational efficiency. What is needed is
an approach that allows to zoom into a specific area which, even
though the number of DOFs treated is different, stays in
equilibrium with the more coarse grained surrounding. The
requirement above has been fulfilled by the recently introduced
AdResS scheme \cite{Praprotnik:2005:4,Praprotnik:2006},
but is more general as the above examples show. In computational
terms such a simulation translates into a scheme of changing the
number of DOFs on demand and on the fly in a selected region.
While examples of robust numerical algorithms have been
given\cite{Praprotnik:2005:4,Praprotnik:2006,Abrams:2005} the underlying concept still
requires to be put into a solid, rigorous theoretical framework,
making it applicable also for the other schemes mentioned before.
This is the subject of the present Letter. The continuous
transition from a less to a more coarse grained description (and
vice-versa) is explained in terms of a ''geometrically induced
first order phase transition'', where the similarity with a
standard phase transition is further put into the context of
non-integer dimensions of the phase space as DOFs are slowly switched
off/on. In this way the present theory leads to the result
of generalizing the equipartition theorem to non-integer
dimensions and shows how to obtain relevant thermodynamic
quantities within a continuous variable resolution of phase space.
In this sense, the original problem of high dimensional systems,
multiscale in nature, can be considerably simplified on the basis
of a general and rigorous statistical mechanics framework. Because
of the generality of the theoretical argument, this should be
applicable to other multiscale simulation approaches.

\begin{figure}[!ht]
\includegraphics[width=7.5cm]{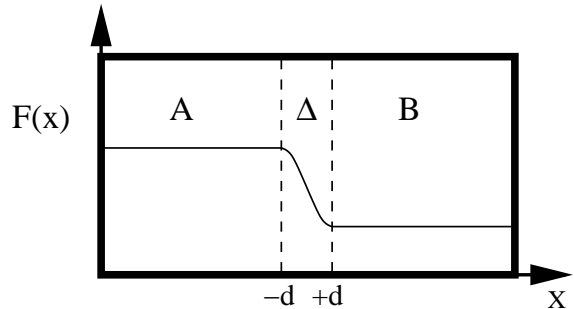}
\caption{\label{free-en} The free energy $F$ as a function of $x$.
  $A$ and $B$ are the regions with high and low levels of detail,
  respectively, while $\Delta$ is a transition regime. The constant
values (minima) of $F_{A}$ and $F_{B}$ are arbitrary and do not play any role
in the following treatment.}
\end{figure}

Without loss of any generality let us assume a system of molecules
in a volume $V$, modeled on a rather coarse-grained level. Now let
us further assume that in a certain subvolume $V'$ a higher
resolution is needed, i.e., to study some function. This is a
typical situation, which one encounters in many systems, e.g.,
proteins or functional molecular assemblies. In statistical terms
this translates into saying that the resolution employed in one
region is lower (or higher) than in the rest. Thus the number of
DOFs of the molecule in such a region is lower (or higher). For
simplicity, we divide the volume $V$ into two parts $A$ and $B$.
In region $A$, each molecule is characterized by $n_{A}$ DOFs, and
in region $B$ by $n_{B}$. For example, in region $A$ one has
higher resolution and a molecule can be considered as a collection
of atoms linked by springs while in $B$ the molecule consists only
of its center of mass and a spherical excluded volume as in our previous
numerical example\cite{Praprotnik:2005:4,Praprotnik:2006}. The
natural question now is how to reach true thermodynamic
equilibrium between the two regions with the same overall
structure of the system on both sides? At this point we assume,
tested numerically in previous
work\cite{Praprotnik:2005:4,Praprotnik:2006,
Henderson:1974,Lyubartsev:1995, Soper:1996,
Tschop:1998:2,Reith:2003,Izvekov:2004,Izvekov:2005,
Abrams:2005} that for a state point $(\rho,T)$ it is in general
possible to reduce the many body potential of the higher
resolution representation into a dimensionally reduced effective
potential. The latter, when applied to a system composed
exclusively of molecules with the lower resolution, reproduces the
statistical properties of a system composed exclusively of highly
resolved molecules, when analyzed accordingly. However,
treating one overall system with regions of
different resolution, requires a special attention. The problem of
the changing DOFs must be addressed in a way that $A$ and $B$
are in thermodynamical equilibrium with each other and,
additionally, the general overall structure is the same. To
address this we visualize the free energy $F$ of the system as a
function of the position $x$. (In order to do so we can divide the
system into "large enough" equal slabs). The free energy is a
thermodynamic potential and hence in thermodynamic equilibrium
$F(x)=F_{A}$, constant in region $A$, and $F(x)=F_{B}$, constant
in region $B$. $F_{A}$ and $F_{B}$ are in general different since
$F$ is an extensive quantity and $n_{A}\neq n_{B}$\cite{Praprotnik:2006:3}.
Note, however, that despite this free energy difference, which
stems exclusively from the different levels of a molecular
representation in regions $A$ and $B$, the chemical potentials
must be equal in both regions. This guarantees that the molecules
experience no spurious driving force which would pull them from
one region into the other due to the choice of the level of
resolution. Figure \ref{free-en} sketches a "typical" free energy
profile across the system.
We focus now on the transition regime $\Delta$ between the two subsystems, i.e.,
$-d\le x\le +d$, where the points $-d$ and $+d$ denote the
boundaries of $\Delta$ with the regions $A$ and $B$, respectively, see
figure \ref{free-en}. In $\Delta$ we gradually change the level of resolution and
consequently the value of $F(x)$.
The width of $\Delta$ is set by the range of the effective pair
potential between molecules\cite{Praprotnik:2005:4}.
Our system is in equilibrium, which implies that at the boundaries
\begin{equation}
 \lim_{x \to d^{-}}\frac{\partial F_{A}(x)}{\partial x}=\lim_{x \to d^{+}}\frac{\partial F_{B}(x)}{\partial x}=0.
\label{eq1}
\end{equation}
If this condition did not hold, a molecule would 'see' a free
energy gradient along $x$ within the same level of resolution
leading to a drift along the $x$ axis. Next, let us write
$\frac{\partial F_{A}}{\partial x}=\frac{\partial F_{A}}{\partial
N_{A}}\frac{\partial N_{A}}{\partial n_{A}}\frac{\partial
n_{A}}{\partial x}$ and the same for $B$, $\frac{\partial
F_{B}}{\partial x}=\frac{\partial F_{B}}{\partial
N_{B}}\frac{\partial N_{B}}{\partial n_{B}}\frac{\partial
n_{B}}{\partial x}$. Here $N_{A}$ and $N_{B}$ are the numbers of
molecules in $A$ and $B$, respectively. $ {\partial
N_{A}}/{\partial n_{A}}$ and ${\partial N_{B}}/{\partial n_{B}}$
are two non-zero constants, while ${\partial F_{A}}/{\partial
N_{A}}=\mu_{A}$ and ${\partial F_{B}}/{\partial N_{B}}=\mu_{B}$,
where $\mu$ is the chemical potential. Note that due to
equilibrium $\mu_{A}=\mu_{B}\ne 0$. The condition of Eq.\ref{eq1}
is hence reduced to:
\begin{equation}
\lim_{x \to d^{-}}\frac{\partial n_{A}(x)}{\partial x}=\lim_{x \to d^{+}}\frac{\partial n_{B}(x)}{\partial x}=0.
\label{eq2}
\end{equation}
Thus, we can formally describe the switching on and off of a given
DOF via a weighting function $w(x)$ such that $w(x)=1; \forall x
\in A$ and $w(x)=0; \forall x \in B$, with zero slope at the
boundaries of $\Delta$. In accordance with Eq.\ref{eq2}, this
requires $w(x)$ to be continuous up to the first derivative and
monotonically goes from the value one to zero in the region
$\Delta$. The latter requirement reflects the fact that we would
like to switch gradually on-the-fly without extra equilibration
from more DOFs to less (or vice-versa)\cite{Praprotnik:2006:2}.
The switching procedure implies that in the transition regime,
where $0\le w(x)\le 1$, we deal with fractional DOFs, i.e., by
switching on/off a DOF we continously change the dimensionality of
the phase space. To rigorously describe the fractional phase space
we resort here to fractional
calculus\cite{Nonnenmacher:1990,Hilfer:2000,
Cottrill:2001,Tarasov:2004,Tarasov:2005}. According to
Refs.\cite{Cottrill:2001,Tarasov:2004,Tarasov:2005} and to the
formula for dimensional regularization \cite{Collins:1984}, the
infinitesimal volume element of the fractional configurational
space is defined as
$dV_\alpha=d^{\alpha}x\,\Gamma(\alpha/2)/(\pi^{\alpha/2}\Gamma(\alpha))
=|x|^{\alpha-1}dx/\Gamma(\alpha)=dx^\alpha/(\alpha\Gamma(\alpha))$
where the positive real parameter $\alpha$ denotes the order of
the fractional coordinate differential and $\Gamma$ is the gamma
function. To make the connection with the switching on/off of DOFs
we consider in our case each $\alpha$ as a value that $w(x)$ can
take and apply the formalism to each DOF separately. Hence, we
have to go beyond Refs. \cite{Tarasov:2004, Tarasov:2005} so that
each DOF can take its own value of parameter $\alpha$ according to
the level to which the particular DOF is switched on. Such a
formulation leads to an interpretation of the switching as
a ''geometrically induced phase transition''. Here, we deal with a
representation-driven geometrical transition, which has formal
similarities with the standard concept of a first-order phase
transition\cite{Lynden-Bell:1995}. In fact, while in a standard
first-order phase transition we have a latent heat, e.g. of
solidification, in this case we have a latent heat due
to the fact that the free energy density becomes position
dependent and we need to furnish or remove ''latent'' heat from an
external bath to compensate the free energy gradient
due to changing the number of DOFs in $\Delta$.

\begin{figure}[!ht]
\includegraphics[width=6cm]{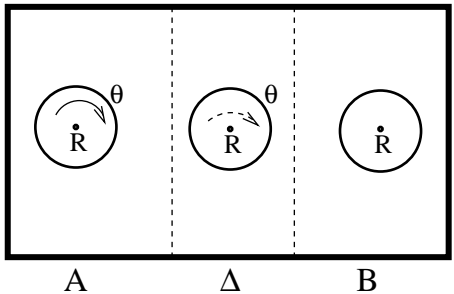}
\caption{\label{kin-en}The molecular resolution of a simple 2-d
  circular molecule in the high and low resolution regions $A$ and $B$
  and the transition region $\Delta$, respectively.}
\end{figure}

To explain this concept in more detail we define the temperature
in $A$ in a usual way as $T_{A}={2\left <K_{A}\right>}/{n_{A}}$,
in $B$, $T_{B}={2\left <K_{B}\right>}/{n_{B}}$, where
$\left<K_{A}\right>$ and $\left<K_{B}\right>$ are the average
kinetic energies of a molecule in regions $A$ and $B$,
respectively; since we have equilibrium, $T_{A}=T_{B}$. In a
similar way we would also like to define the temperature in the
interface region $\Delta$ as
$T_{\Delta}={2\left<K_{\Delta}\right>}/{n_{\Delta}}$ with
$\left<K_{\Delta}\right>$ and $n_{\Delta}$ being the average kinetic
energy and the number of DOFs of a molecule in a given slab $x$
in $\Delta$, respectively.
 However, for such a definition we first need to determine how
 $K_{\Delta}$ and $n_{\Delta}$ scale with $w(x)$.
Let us demonstrate this with a simple example. In figure
\ref{kin-en} we have a disc representing a 2-d molecule with three
DOFs in the high resolution region $A$, i.e., two translational
DOFs of the center of mass ${\bf R}=(R_x,R_y)$ and one rotational
DOF around the center of mass characterized by angle $\Theta$, the
transition representation of the molecule in the region $\Delta$,
and the coarse grained molecule in the region $B$ with only two
translational DOFs.
The kinetic energy in the region $A$, setting the mass, the
molecule's radius, and the Boltzmann constant $m=r=k_B=1$, is:
$K_{A}=[{\dot{\bf R}}^{2}+p_\Theta^{2}]/2=
[\dot{R_x}^{2}+\dot{R_y}^{2}+\dot{\Theta}^{2}]/2$, and in region
$B$: $K_{B}=\dot{{\bf R}}^{2}/2=[\dot{R_x}^{2}+\dot{R_y}^{2}]/2$.
Here $p_{\Theta}$ denotes the angular momentum. According to the equipartition theorem each
full quadratic DOF, i.e., $R_x$, $R_y$, and $\Theta$, contributes
to the kinetic energy with an amount of $T/2$. Hence,
$T_{A}={2<K_{A}>}/{3}$, $T_{B}={2<K_{B}>}/{2}$. In the region
$\Delta$ however, $\Theta$ is not a full DOF, because its weight
varies between zero and one. This has to be considered, when
calculating the local temperature.
To determine such a contribution one should account for the
following: Let us consider a given value $w(x)=\alpha$ and use
$\alpha$ as a variable parameter. The instantaneous kinetic energy
of a fractional DOF should gradually vanish as the DOF is slowly
switched off (or vice-versa), thus in the region $\Delta$, our
ansatz for the kinetic energy associated to $\Theta$ is
$f(\alpha)p_{\Theta}^{2}/2$, where $f(\alpha)$ is a monotonic
function in $\Delta$ with $f(1)=1;f(0)=0$. Apart from these
requirements we do not need to specify the exact form of
$f(\alpha)$ at this point. Accordingly, the kinetic energy in the
position with the coordinate $x$ in the region $\Delta$ is
$K_{\Delta}=[\dot{R_x}^{2}+\dot{R_y}^{2}+f(\alpha)p_\Theta^{2}]/2$. For the
fractional quadratic DOF $\Theta$ we can then write the partition
function as:
\begin{multline}
 \exp(-\beta F_{\alpha})=C\int\exp(-\beta f(\alpha)p_{\Theta}^{2}/2)\,dV_\alpha=\\=2C\int_{0}^{\infty}\exp(-\beta f(\alpha)p_{\Theta}^{2}/2)\,|p_{\Theta}|^{\alpha-1}\frac{dp_{\Theta}}{\Gamma(\alpha)}=\\=\frac{2^{\alpha/2}C\Gamma(\alpha/2)}{\Gamma(\alpha)}f(\alpha)^{-\alpha/2}\beta^{-\alpha/2}\sim\beta^{-\alpha/2},
\label{eq3}
\end{multline}
where $C$ is a normalization constant, $\beta=1/T$, and $F_\alpha$
the free energy associated with the fractional DOF $\Theta$,
respectively\cite{Feynman:1965}. The consequence of Eq.\ref{eq3}
is the fractional analog of the equipartition theorem:
\begin{equation}
 \left<K_\alpha\right>=\frac{d(\beta F_\alpha)}{d\beta}=\frac{\alpha}{2\beta}=\frac{\alpha T}2,
\label{eq4}
\end{equation}
where  $\left<K_\alpha\right>$ is the average kinetic energy per
fractional quadratic DOF with the weight $\alpha$. Thus, for
$\alpha=0,1$ we obtain the correct limits in the coarse-grained and
fully resolved regimes, respectively, with the correct contributions
to the kinetic energy. Furthermore, we have
$T_{\Delta}=2\left<K_{\Delta}\right>/n_{\Delta}=(2+\alpha)
T/n_{\Delta}$. To satisfy the equilibrium condition:
$T_A=T_B=T_{\Delta}=T$ we must set $n_{\Delta}=2+\alpha$, which is
in accordance with the ``intuitive'' definition in Ref.
\cite{Praprotnik:2005:4}. The number of quadratic DOFs and the
average kinetic energy thus scale linearly with $w(x)$.
Eq.\ref{eq4} also tells us that, although the equipartition is
independent of the specific choice of $f(\alpha)$, since the
average kinetic energy scales as $\alpha$ also the instantaneous
one should scale in the same way. This means that
$f(\alpha)=\alpha$ and it is determined by the fractional
character of the phase space. Note that for non-quadratic DOFs the
functional form of $f(\alpha)$ is generally more
complicated\cite{Deserno:2006}, however not needed for the present
purpose.

Based on this theoretical considerations we have recently derived
an efficient particle-based MD simulation
scheme\cite{Praprotnik:2005:4,Praprotnik:2006}. For intermolecular
force calculation we use an interpolation formula for the force
acting between centers of mass of given molecules $\alpha$ and
$\beta$:
\begin{equation}
 {\bf F}_{\alpha\beta}=w(x_\alpha)w(x_\beta){\bf
 F}_{\alpha\beta}^{atom}+[1-w(x_\alpha)w(x_\beta)]{\bf
 F}_{\alpha\beta}^{cg}\label{eq5}
\end{equation}
where $x_\alpha$ and $x_\beta$ are the center of mass coordinates
of the molecules $\alpha$ and $\beta$, respectively, ${\bf
F}_{\alpha\beta}^{atom}$ is the sum of all pair atom interactions
between explicit atoms of molecule $\alpha$ and explicit atoms of
molecule $\beta$ and ${\bf F}_{\alpha\beta}^{cg}$ is the total
force between the centers of mass of the respective two
molecules\cite{Praprotnik:2005:4,Praprotnik:2006}. This ansatz
satisfies Newton's Third Law and takes into account the transfer
of the turned off explicit DOF onto the molecular center of mass\cite{Praprotnik:2006:4}.
Each time a given molecule crosses a boundary between different
  regimes it gains or looses (depending on whether it leaves or enters the region $B$)
its equilibrated rotational DOF while retaining its linear momentum.
By extension of the equipartition theorem to fractional DOFs we are
able to define the means to supply the latent heat, which is required
or removed for the transition from coarse-grained to high resolution or
vice-versa. Since this generalized equipartition theorem also applies
to the fractional quadratic DOFs standard thermostats based on the
fluctuation-dissipation theorem are applicable.

In conclusion, we provided the statistical mechanics foundation
for an efficient computational scheme that concurrently couples
different length scales via different levels of resolution, i.e.,
atomistic and mesoscopic length scales, by adapting the resolution
on demand. The transition region is well defined by the here
introduced generalization of the equipartition theorem for
fractional dimension of phase space. While it directly applies to
a scheme recently tested by the
authors\cite{Praprotnik:2005:4,Praprotnik:2006} it in the same way
should also provide the general theoretical framework to extend
other commonly used schemes, such as
\cite{Csanyi:2004,Flekkoy:2000, Delgado:2003} towards a truly
adaptive multiscale simulation scheme.

We thank M.~Deserno, D.~Andrienko, and L.~Ghiringhelli for helpful
comments on the manuscript. This work is supported in part by the
Volkswagen foundation. M.~P. acknowledges the support of the
Ministry of Higher Education, Science and Technology of Slovenia
under grant No. P1-0002.




\begin{thebibliography}{10}
\bibitem{Praprotnik:2006:3}
The fact that $F_A$ and $F_B$ need not to be equal illustrates the limitations
  of the analogy to phase transitions, where the free energy at the transition
  point is the same in the two phases.
\bibitem{Praprotnik:2006:2}
Eq.2 expresses the fact that we cannot change $n_A$ and $n_B$ otherwise than by
  changing $N_A$ and $N_B$, respectively. In contrast, we also change the
  number of DOFs in $\Delta$ by changing the number of DOF per molecule with
  the weighting function $w(x)$.
\bibitem{Praprotnik:2006:4}
One could, in principle, also define the coupling scheme using the respective
  potentials instead of forces as
  $U_{\alpha\beta}=w(x_\alpha)w(x_\beta)U_{\alpha\beta}^{atom}+[1-w(x_\alpha)w%
(x_\beta)] U_{\alpha\beta}^{cg}$. Note that such a linear combination depends
  on the absolute positions of the particles and not only on their relative
  distances. A force defined as a gradient of this potential would not obey
  Newton's Third Law unless $w(x)=const$.
\end{thebibliography}

\end{document}